\RequirePackage{fix-cm}
\documentclass[twocolumn]{svjour3}          
\smartqed  
\usepackage{graphicx} 
\usepackage{epstopdf,epsfig,color,amssymb,amsfonts,amsmath,fancybox} 
 \usepackage{mathptmx}      
\usepackage{latexsym}
 \journalname{Systems and Synthetic Biology}
\begin{document}

\title{Systems biology beyond degree, hubs and scale-free networks: the case for multiple metrics in complex networks}



\author{Soumen Roy}


\institute{Soumen Roy \at Bose Institute, 93/1 Acharya PC Roy Road, Kolkata 700 009 India\\
              \email{soumen@boseinst.ernet.in}}

\date{Received: \today ~ / Accepted: date}

\maketitle

\begin{abstract}
Modeling and topological analysis of networks in biological and other complex systems, must venture beyond the limited consideration of very few network metrics like degree, betweenness or assortativity. A proper identification of informative and redundant entities from many different metrics, using  recently demonstrated techniques, is essential. A holistic comparison of networks  and growth models is best achieved only with the use of such methods. 
\end{abstract}

Network theory 
(Albert et al. 2002, Newman 2010) 
plays an important role in Systems Biology. Complex network literature is replete with  discussions  about networks bearing knowledge of function, signatures of complexity and  information about ``emergent properties" of the system being encoded in their topology. It is therefore quite natural to assume that a comprehensive study of a significant number of network metrics would convey  a lot of information about the system. Interestingly however, most papers in literature analyze at most two or three  metrics at a time. Thus arises a very relevant yet unanswered question - do these few handpicked network metrics convey most of the knowledge that could have been known about the network? 

The  idea that the network topology can be a major determinant of function (or dysfunction) has been studied in considerable detail.  
The relation between the topological properties of network nodes (genes, proteins) and functional essentiality is well known in interaction networks  (Albert et al. 2000,  Jeong et al. 2001). 

In metabolic networks, long before the advent of the complex networks era, extensive modeling had been done using steady-state flux balance approaches (Varma et al. 19-94) 
via  methods like Flux Balance Analysis (FBA) (Edwards et al. 2000), 
Minimization of Metabolic Adjustment (MO-MA) (Segre et al. 2002), 
and Elementary Mode Analysis (EMA) (Steling et al. 2002). 
Nevertheless,  topological analysis has often yielded novel and valuable insight, in metabol-ic networks.  For example, new parameters  like synthetic accessibility (SA) have exhibited sufficient power in predicting the viability of knockout strains with accuracy comparable to approches using biochemical parameters (like FBA etc.) on large, unbiased mutant data sets (Wunderlich et al. 2006). 
This is especially remarkable since determining SA does not require the knowledge of stoichiometry or maximal uptake rates for metabolic and transport reactions which might be necessary in FBA, MOMA and EMA. Also, it can be rapidly computed for a given network and has no adjustable parameters.

Degree or the number of connections a node has with other nodes in the network and sometimes also with itself, is the most common topological metric in networks. It is perhaps hard to find a paper in complex networks which does not mention degree.  Degree distributions are generally well-studied for almost all systems. Unfortunately, there is still a  trend of labeling networks possessing heavy-tailed degree distributions as scale-free networks, i.e., networks having a power-law degree distribution. This is widespread, in spite of a  reliable statistical machinery for proper identification of scale-free networks (Clauset et al., 2009). 

Power-laws have a special place in statistical physics, and hence the activity around ``scale-free networks" in phys-ics literature is somewhat understandable . However, one is at a loss to comprehend  as to why ``scale-free networks" are overemphasized in the biological networks community. Especially, when many degree distributions could be fit equally well or perhaps even better by other distributions. Irrespective of whether they obey a power-law or not, all heavy-tailed degree distributions have at least one thing in common: hubs or high degree nodes in the  network. The overreaching engagement with ``hubs"  seems to stem from the apparent conclusion that removal of these could cause massive damage to the network.  However, it was shown quite sometime ago by means of the ``S-metric", that even with a scaling degree sequence,  extremely important networks like the internet could be structurally robust and functionally stable (Doyle et al. 2005). 
Thus, the removal of hubs might not necessarily have a catastrophic but merely a local effect. 

\begin{figure}
\includegraphics[width=\columnwidth, height=8cm, angle=0]{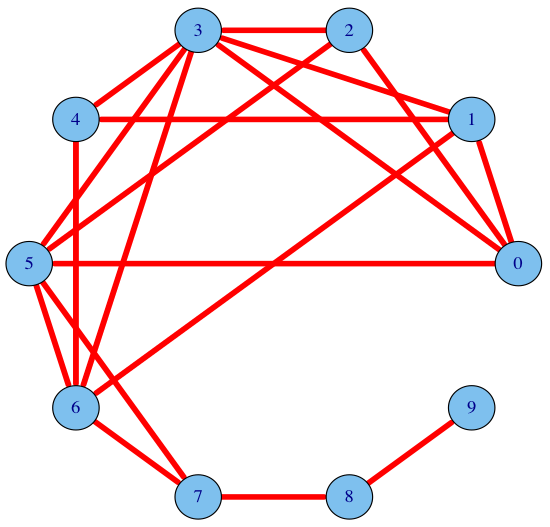}
\caption{An elementary analysis would reveal that targeting a high-betweenness node like 7 over  a `hub' (like 3) would cause much more damage to the network.}
\label{fig:betweenness}       
\end{figure}
There have been a number of works showing that hubs are not always the most  important nodes. Social scientists have known this for a long time  via the analysis of graphs like the ``Krackhardt kite graph" shown in Fig.~\ref{fig:betweenness} (Wasserman et al. 1994).  
One of the most important  properties of a network node is ``betweenness centrality", which measures the fraction of all shortest paths passing through that node (Freeman 1977). 
In the world air transportation network, the common perception would  probably be that most shortest flights between any two airports are likely to pass through cities like London and New York. Actual analysis showed that many of the shortest paths did not pass through $60\%$ of the $25$  most connected airports. Instead many of them passed through airports like Anchorage and Port Moresby  (Guimera et al. 2005). 
Of course carrying forward this treatment from an unweighted to a  weighted network might change the results somewhat but their general significance is not lost. 
It is also known that maximum damage would be done to the US airline network if the airports are targeted by betweenness rather than hubness (Wuellner et al. 2010). 
A number of papers using biological networks have found important results using betweenness (Dunn et al. 2005, Hahn et al. 2005, Joy et al. 2005, Hegde et al. 2008, Liu et al. 2009). 

Assortative mixing (Newman 2002), i.e. whether high-degree nodes are connected to other high-degree nodes in a network, is also known to be an important consideration in a number of biological networks  (Bagler et al. 2007, Pechenick et al. 2012). 
While it was earlier thought that all biological networks are disassortative, it has been subsequently found that protein contact networks could be assortative (Bagler et al. 2007). 

It should be mentioned here that various measures from spectral graph theory are known to shed valuable insight in graphs and have also been studied extensively in biological networks (Banerjee et al. 2009, Perkins  and Langston 2009).  

Thus, it is abundantly clear that in some circumstances degree is an important metric; while in some others it might be betweenness or assortativity and so on. This naturally begs the question as to how one can identify  which metrics are important in a given scenario and which ones are redundant. 

In recent literature, an appropriate quantitative framework has been proposed to address this issue this by incorporating multiple network metrics and higher moments of some of these  (Filkov et al. 2009, Roy et al. 2009).  
These papers considered a significant number of metrics, including higher moments of metric distributions, wherever appropriate. Many distributions are often (albeit not always), quantified by their first few moments. For example, distributions of metrics like degree, betweenness, geodesic or clustering might carry important information about the system and should be studied in depth whenever possible.   Methods from data mining such as clustering and statistical dimension reduction techniques like  Principal Component Analysis (PCA) (Jolliffe 2002) can then be utilized for the unambiguous identification of informative and redundant network metrics. 
The results obtained by this treatment clearly demonstrate that it is not just the degree or betweenness or some other metric which is important. Most of the meaningful information is actually carried by a linear combination of some metrics and/or the higher moments of a few metrics (Filkov et al. 2009, Roy et al. 2009). The essence of usage of these techniques is outlined below. 

A heatmap is a typical tool in clustering which is extensively used across the sciences. In a heatmap, the rows and columns are arranged so that, the most correlated metrics are placed closest to each other due to the hierarchical clustering used. Heatmaps allow us to identify clusters of {\em similar} network attributes by detecting blocks of squares along the diagonal. A limited amount of clustering along the diagonal would imply that most of the network metrics  chosen are effectively independent and could be informative for our analysis. On the other hand, sizable blocks along the diagonal would denote redundancy. 

Well known statistical dimension-reduction techniques, like Principal Components Analysis allow for a comprehensive comparison across many metrics in networks. The essential idea at the heart of PCA is to  ensure that when high dimensional data is projected to a lower dimension, the maximum variance is retained. PCA enables the projection of an n - dimensional dataset onto an equi-dimensional space,  such that the ``new axes" (in other words, the principal components) are orthogonal. These principal components are actually linear combinations of the original dimensional variables, such that the first $d$ axes, where $d \le n$ retains the maximal variance of the original data set. 

The power of these methods lies in the fact that they can also be used for comparing network growth models among themselves and how individual models fare with respect to real world networks.  (Filkov et al. 2009). 

One might wonder if the consideration of the first few moments of a distribution is a mere book keeping exercise.  That they are indeed informative is reflected by the emphatic presence of a number of higher moments of metric distributions in the first few principal components of analyzed data and/or models (Filkov et al. 2009, Agarwal et al. 2010, Vilar et al. 2010, Bounova et al. 2012). Again, metabolic networks apparently bear strong signatures of organism phenotypes in some  of the higher moments of their network metrics (Roy et al. 2009).  


In conclusion, the above discussion hopefully establishes the importance of the fact that it is only proper that topological analysis and modeling of networks in systems biology and complex networks should venture beyond the treatment of betweenness, degree, hubs,  scale-free networks etc. and instead focus on multiple metrics in networks. 


 
\end{document}